# Distributed Active Noise Control System Based on a Block Diffusion F*x*LMS Algorithm with Bidirectional Communication

Tianyou Li, Hongji Duan, Sipei Zhao, *Member, IEEE*,

Jing Lu, *Member, IEEE*, and Ian S. Burnett, *Senior Member, IEEE*

*Abstract*—Recently, distributed active noise control systems based on diffusion adaptation have attracted significant research interest due to their balance between computational complexity and stability compared to conventional centralized and decentralized adaptation schemes. However, the existing diffusion F*x*LMS algorithm employs node-specific adaptation and neighborhood-wide combination, and assumes that the control filters of neighbor nodes are similar to each other. This assumption is not true in practical applications, and it leads to inferior performance to the centralized controller approach. In contrast, this paper proposes a Block Diffusion F*x*LMS algorithm with bidirectional communication, which uses neighborhood-wide adaptation and node-specific combination to update the control filters. Simulation results validate that the proposed algorithm converges to the solution of the centralized controller with reduced computational burden.

*Index Terms*—Multichannel ANC, Distributed Diffusion F*x*LMS, computational complexity

## I. INTRODUCTION

CONVENTIONAL multichannel active noise control (ANC) systems based on centralized adaptation suffer from high computational burden [1, 2]. In contrast, decentralized adaptation regulates each channel independently by ignoring acoustic inter-channel coupling [3], reducing computational complexity but introducing instability issues [4]. Over the past decade, a distributed ANC system based on Wireless Acoustic Sensor and Actuator Networks (WASANs) has been developed in several papers to balance computational load and convergence behavior [5, 6]. The distributed ANC system consists of multiple nodes, each of which includes a processor, one or more error microphones and one or more secondary loudspeakers. Incremental [7-9] and diffusion [10-12] adaptations are the most commonly used collaboration strategies in distributed network, but the former is sensitive to link failures between nodes [13, 14]. Therefore, this letter focuses on the diffusion adaption schemes.

Generally, the loudspeakers and microphones of each node are placed at different physical locations relative to the primary noise source such that the optimal control filters (Wiener solution) for each node vary from one another [15]; this results in the multitask estimation of the control filters in multichannel ANC systems. Therefore, the Multitask Diffusion F*x*LMS (MDF*x*LMS) algorithm has been explored to update each node's control filter based on its own error signal first, and then spatially average the control filters over the node's neighborhood in each iteration [16]. Effectively, this spatial regularization forces the control filters in the neighborhood to be more similar to each other, which may introduce bias in the converged solution [17, 18] and thus the approach cannot achieve the same performance as the centralized multichannel F*x*LMS algorithm [13, 19]. To alleviate such performance deterioration, an improved algorithm based on the Variable Spatial Regularization (MDF*x*LMS-VSR) has been proposed recently [20, 21]. However, the VSR strategy can only alleviate (and not eliminate) the bias and requires the Wiener solution as *a priori* information, which is usually unavailable in practice.

To overcome this problem, this letter proposes a new Block Diffusion F*x*LMS algorithm with Bidirectional Communication (BDF*x*LMS-BC), which employs neighborhood-wide adaptation and node-specific combination. The algorithm is derived in detail and simulation results of a 10-channel ANC system demonstrate that the proposed algorithm eliminates the bias in the MDF*x*LMS algorithm, converging to the solution of the centralized multichannel F*x*LMS algorithm.

Manuscript received —, 2022; revised —, 2022; accepted —, 2022. Date of publication —, 2022; date of current version —, 2022. The National Science Foundation of China supported this work with grant number 12274221. The associate editor coordinating the review of this manuscript and approving it for publication was —. (Corresponding author: Jing Lu)

T. Li, H. Duan and J. Lu are with the Key Laboratory of Modern Acoustics, Nanjing University, Nanjing 210008, China (email: tianyou.li@smail.nju.edu.cn; duan@smail.nju.edu.cn; lujing@nju.edu.cn )

S. Zhao and I. S. Burnett are with the Center for Audio, Acoustics and Vibration, Faculty of Engineering and IT, University of Technology Sydney, Ultimo, NSW 2007 Australia (email: sipei.zhao@uts.edu.au; Ian.Burnett@uts.edu.au )

Digital Object Identifier —

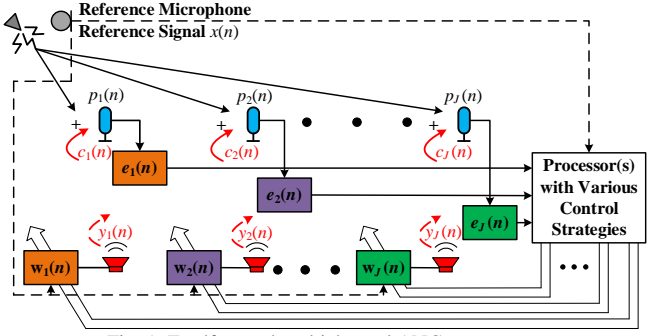

Fig. 1. Feedforward multichannel ANC system.

## II. EXISTING MULTICHANNEL ANC ALGORITHMS

A feedforward multichannel ANC system (Fig. 1) consists of a single reference microphone to capture the reference sound signal, $J$ loudspeakers to produce the control sound, and $J$ error microphones to measure the residual noise. The error signals $\mathbf{e}(n) = [e_1(n) \ldots e_J(n)]^T$ can be written as

$$\mathbf{e}(n) = \mathbf{p}(n) + \mathbf{c}(n), \quad (1)$$

where $\mathbf{p}(n) = [p_1(n) \ldots p_J(n)]^T$ and $\mathbf{c}(n) = [c_1(n) \ldots c_J(n)]^T$ are the primary noise and control sound at the error microphones, respectively, and the superscript $^T$ denotes transposition. The control sound generated by all the loudspeakers are

$$\mathbf{c}(n) = \begin{bmatrix} c_1(n) \\ c_2(n) \\ \vdots \\ c_J(n) \end{bmatrix} = \begin{bmatrix} \mathbf{h}_{11}^T & \mathbf{h}_{12}^T & \cdots & \mathbf{h}_{1J}^T \\ \mathbf{h}_{21}^T & \mathbf{h}_{22}^T & \cdots & \mathbf{h}_{2J}^T \\ \vdots & \vdots & \ddots & \vdots \\ \mathbf{h}_{J1}^T & \mathbf{h}_{J2}^T & \cdots & \mathbf{h}_{JJ}^T \end{bmatrix} \begin{bmatrix} \mathbf{y}_1(n) \\ \mathbf{y}_2(n) \\ \vdots \\ \mathbf{y}_J(n) \end{bmatrix}, \quad (2)$$

where $\mathbf{h}_{jk} = [h_{jk,0} \ldots h_{jk,H-1}]^T$ denote the secondary path between the $k^{\text{th}}$ loudspeaker and the $j^{\text{th}}$ error microphone. The output signal of the $j^{\text{th}}$ loudspeakers, $\mathbf{y}_j(n) = [y_j(n) \ldots y_j(n-H+1)]^T$, can be obtained by filtering the reference signal with the corresponding control filter, i.e.,

$$\mathbf{y}_j(n) = \left[\mathbf{x}(n)\,\mathbf{x}(n-1)\cdots\mathbf{x}(n-J+1)\right]^T \mathbf{w}_j(n), \quad (3)$$

where $\mathbf{w}_j(n) = [\mathbf{w}_{j,0}(n) \ldots \mathbf{w}_{j,I-1}(n)]^T$ is the $j^{\text{th}}$ $I$-tap control filter and $\mathbf{x}(n) = [x(n) \ldots x(n-I+1)]^T$ is the reference signal vector. It is assumed that the control filters change slowly according to the conventional multichannel ANC theory [22, 23]. Substituting (3) and (2) into (1) yields

$$\mathbf{e}(n) = \mathbf{p}(n) + \mathbf{F}(n)\mathbf{w}(n), \quad (4)$$

where $\mathbf{w}(n) = \left[\mathbf{w}_1^T(n) \cdots \mathbf{w}_J^T(n)\right]^T$ is the global control filters and

$$\mathbf{F}(n) = \begin{bmatrix} \mathbf{F}_{1,1}^T(n) & \mathbf{F}_{1,2}^T(n) & \cdots & \mathbf{F}_{1,J}^T(n) \\ \mathbf{F}_{2,1}^T(n) & \mathbf{F}_{2,2}^T(n) & \cdots & \mathbf{F}_{2,J}^T(n) \\ \vdots & \vdots & \ddots & \vdots \\ \mathbf{F}_{J,1}^T(n) & \mathbf{F}_{J,2}^T(n) & \cdots & \mathbf{F}_{J,J}^T(n) \end{bmatrix} \quad (5)$$

is the global filtered-$x$ signal matrix, where $\mathbf{F}_{j,k}(n) = \left[\mathbf{x}(n)\,\mathbf{x}(n-1)\cdots\mathbf{x}(n-H+1)\right]\hat{\mathbf{h}}_{jk}$, with $\hat{\mathbf{h}}_{jk}$ being the secondary path model that can be obtained by offline estimation methods [22]. The conventional Centralized F$x$LMS (CF$x$LMS) and Decentralized F$x$LMS (DCF$x$LMS) algorithms aim to minimize the squared global error signals [2] and the squared single-channel error signal [3, 4], respectively. Using the stochastic gradient descent method [24], the normalized update equations for the CF$x$LMS [2] and DCF$x$LMS [3, 4] are obtained as

$$\mathbf{w}(n+1) = \mathbf{w}(n) - \mu^C \mathbf{F}_{nor}(n)\mathbf{e}(n) \quad (6)$$

and

$$\mathbf{w}_j(n+1) = \mathbf{w}_j(n) - \mu_j^D \frac{\mathbf{F}_{j,j}(n)}{\left\|\mathbf{F}_{j,j}(n)\right\|^2} e_j(n), \quad (7)$$

respectively, where $\mu^C$ and $\mu_j^D$ are the step sizes and $\|\bullet\|^2$ denotes the squared Euclidean norm. $\mathbf{F}_{nor}(n)$ in (6) is a block matrix composed of $\mathbf{F}_{s,t}^T(n) / \left\|\mathbf{F}_{s,t}^T(n)\right\|^2$ $(1 \leq s, t \leq J)$.

The multitask diffusion adaptation shares information between neighbor nodes to reduce the instability risk, hence its cost function includes the squared Euclidean distance between the control filters within the neighborhood as an additional spatial regularization term [17, 18]. The distributed ANC algorithm can be derived by using the Adapt-then-Combine (ATC) diffusion strategy [17, 18]:

$$\tilde{\mathbf{w}}_j(n+1) = \mathbf{w}_j(n) - \mu_j^M \frac{\mathbf{F}_{j,j}(n)}{\left\|\mathbf{F}_{j,j}(n)\right\|^2} e_j(n) \quad (8)$$

and

$$\mathbf{w}_j(n+1) = \sum_{l \in \mathcal{N}_j} c_{jp} \tilde{\mathbf{w}}_j(n+1), \quad \sum_{p \in \mathcal{N}_j} c_{jp} = 1 \quad (9)$$

where $\mu_j^M$ is the step size of the $j^{\text{th}}$ node processor and $\tilde{\mathbf{w}}_j$ is the adaptation result and the index set $\mathcal{N}_j$ denotes the neighborhood of the $j^{\text{th}}$ node, including $j$. The original MDF$x$LMS algorithm based on the time-invariant combination coefficients $\{c_{jp}\}$ assumes similarity between the control filters of different nodes [17, 19]. Meanwhile, the MDF$x$LMS-VSR algorithm utilizes variable combination coefficients based on the *a priori* Wiener solution to improve estimation accuracy and stability. The details of the MDF$x$LMS-VSR algorithm are described in [20, 21].

## III. PROPOSED ALGORITHM

In this section, a new BDF$x$LMS-BC algorithm involving neighborhood-wide adaptation and node-specific combination is proposed. In the adaptation phase, as shown in Step 1 of Fig. 2, each node processor estimates a block vector consisting of the node's own and its neighbor nodes' control filters. In the combination phase, each node updates its own control filter by combining the different estimations of its neighbor nodes (Step 2 of Fig. 2) first. This is termed forward communication, and is followed by a broadcast of the node's updated control filter to its neighbor nodes for the next iteration (Step 3 of Fig. 2). The latter is referred to as the backward communication. This bidirectional communication scheme enables the system to improve stability without sacrificing in performance.

To formulate the BDF$x$LMS-BC algorithm, a block vector for the $j^{\text{th}}$ node processor is constructed as

$$\mathbf{u}_j = col\{\mathbf{w}_p\}, p \in \mathcal{N}_j, \quad (10)$$

where the operator $col\{\bullet\}$ stacks inner control filter arguments on each other according to the element order of the index set $\mathcal{N}_j$. Similarly, the estimate of the block vector $\mathbf{u}_j$ is defined as $\varsigma_j = col\{\tilde{\mathbf{w}}_p^j\}, p \in \mathcal{N}_j$, where the inner control filter $\tilde{\mathbf{w}}_p^j$ denotes the $p^{th}$ control filter estimated by the $j^{th}$ node processor.

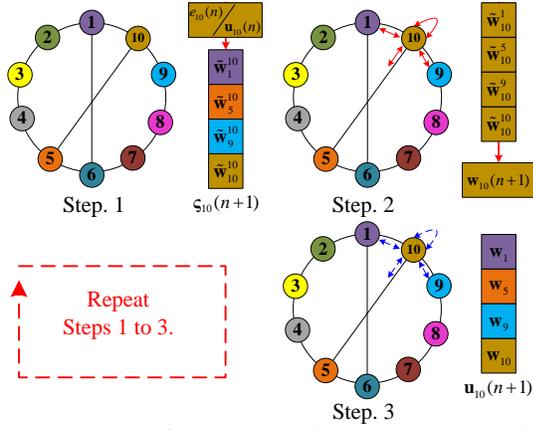

Fig. 2. The iteration of the BDF*x*LMS algorithm with bidirectional communication for a 10-node network.

To measure the deviation between the estimations of the $j^{th}$ control filter and the optimal solution, a block-based regularization term is defined as

$$\Delta_j = \sum_{p \in \mathcal{N}_j} \chi_{jp} \left\| \mathbf{w}_j^{opt} - \tilde{\mathbf{w}}_j^p \right\|^2 = \sum_{p \in \mathcal{N}_j} \chi_{jp} \left\| \mathbf{S}_j^j \mathbf{u}_j^{opt} - \mathbf{S}_j^p \varsigma_p \right\|^2, \quad (11)$$

where $\mathbf{w}_j^{opt}$ denotes the optimal control filter of the $j^{th}$ processor and $\mathbf{u}_j^{opt} = col\{\mathbf{w}_p^{opt}\}, p \in \mathcal{N}_j$ is the optimal block vector. The non-negative coefficients $\{\chi_{jp}\}$ are used to reflect the influence of different neighbor nodes $p \in \mathcal{N}_j$ on the estimation of the $j^{th}$ control filter. The selection matrix $\mathbf{S}_j^p = (\mathbf{s}_j^p)^T \otimes \mathbf{G}_I$ extracts the $j^{th}$ control filter from the $p^{th}$ block vector, $\mathbf{s}_j^p$ is a $|\mathcal{N}_j| \times 1$ column vector with unique non-zero elements at the $s^{th}$ row for $\mathcal{N}_j(s) = p$, $|\mathcal{N}_j|$ denotes the cardinality of $\mathcal{N}_j$, the operator $\otimes$ denotes the Kronecker product and $\mathbf{G}_I$ is the identity matrix of dimension $I$. Based on the block-based regularization, the cost function of the $j^{th}$ node is defined as

$$J_j^B(\mathbf{u}_j) = \mathbb{E}\left[e_j^2(n)\right] + \eta_j \sum_{p \in \mathcal{N}_j} \chi_{jp} \left\| \mathbf{S}_j^j \mathbf{u}_j^{opt} - \mathbf{S}_j^p \varsigma_p \right\|^2 \quad (12)$$

where the spatial parameter $\eta_j$ balances the local error minimization and block-based regularization term, and the unknown $\mathbf{u}_j^{opt}$ can be approximated by $\mathbf{u}_j$ in iteration [25].

Using the ATC strategy [25, 26], the adaptation and combination equation (after the forward communication in Step 2 of Fig. 2) of the $j^{th}$ node processor can be derived as

$$\varsigma_j(n+1) = \mathbf{u}_j(n) - \mu_j^B e_j(n) \frac{\hat{\mathbf{f}}_{F,j}(n)}{\left\| \hat{\mathbf{f}}_{F,j}(n) \right\|^2} \quad (13)$$

and

$$\mathbf{w}_j(n+1) = \tilde{\mathbf{w}}_j^j(n+1) - \mu_j^B \eta_j \sum_{p \in \mathcal{N}_j} \chi_{jp} \left( \mathbf{w}_j(n) - \tilde{\mathbf{w}}_j^p(n) \right), \quad (14)$$

respectively. $\mu_j^B$ is the step size of the $j^{th}$ node processor based on the BDF*x*LMS-BC algorithm. The neighborhood-based filtered reference signal $\hat{\mathbf{f}}_{F,j}(n) = col\{\mathbf{F}_{j,p}(n)\}\left(p \in \mathcal{N}_j\right)$. $\mathbf{w}_j(n)$ and $\tilde{\mathbf{w}}_j^p(n)$ in (14) can be approximated with $\tilde{\mathbf{w}}_j^j(n+1)$ and $\tilde{\mathbf{w}}_j^p(n+1)$, respectively, to account for the latest error signals [17, 27]. Further, Eq. (14) can be simplified as

$$\mathbf{w}_j(n+1) = \sum_{p \in \mathcal{N}_j} c^{jp} \tilde{\mathbf{w}}_j^p(n+1) \quad (15)$$

where the intermediate coefficients $\eta_j$ and $\{\chi_{jp}\}$ have been integrated to the nonnegative combination coefficients $\{c_{jp}\}$ that satisfy $\sum_{p \in \mathcal{N}_j} c^{jp} = 1$ [26]. The latter can be determined by the Uniform [28], Laplacian [29], Metropolis [30] or other weighted combination rules. It is noted that, in Eq. (13), the block vector $\varsigma_j(n+1)$ is updated based on $\mathbf{u}_j(n)$ rather than $\varsigma_j(n)$ from last iteration, hence the backward communication (Step 3 of Fig. 2) is needed to transfer the updated control filters to the neighbor nodes, i.e.,

$$\mathbf{u}_j(n+1) = \mathcal{M}_j \mathbf{w}(n+1) \quad (16)$$

where $\mathcal{M}_j = \mathbf{M}_j \otimes \mathbf{G}_I$ is a $|\mathcal{N}_j|I \times JI$ matrix for the $j^{th}$ node processor, with the $|\mathcal{N}_j| \times J$ matrix $\mathbf{M}_j(s,t) = \begin{cases} 0, \mathcal{N}_j(s) \neq t \\ 1, \mathcal{N}_j(s) = t \end{cases}$.

To summarize, Eqs. (13), (15) and (16) form a complete iteration cycle of the proposed BDF*x*LMS-BC algorithm, corresponding to the three steps in Fig. 2, which are iterated until the system converges.

## IV. SIMULATIONS

A 10-channel ANC system, as illustrated in Fig. 3, and is simulated with free-field impulse responses to compare the performance of the proposed algorithm with other existing algorithms. The radii of the circular error microphone and secondary loudspeaker arrays are 1.0 m and 1.2 m, respectively and the primary noise source is placed 2 m away from the center of the arrays. The primary paths, secondary paths and control filters are modeled as 64, 64 and 260 tap FIR filters, respectively. The output signal of the primary noise source is taken as the ideal reference signal and the signal-to-noise ratio (SNR) of all microphones are set to 30 dB. A 300 Hz tonal signal and a bandpass-filtered (100 Hz – 1500 Hz) white Gaussian noise are used as the primary noise signals, respectively.

The Metropolis rule is used to determine the fixed combination parameters for the MDF*x*LMS and BDF*x*LMS algorithms and the initial combination parameters for VSR strategy. The communication links for all the diffusion algorithms are shown in Fig. 2. The step sizes for the centralized and diffusion algorithms are selected to be as large as possible to achieve the fastest convergence while maintaining the best noise reduction performance. The step sizes of the CF*x*LMS algorithm for the tonal signal and broadband noise are 2.8 and 1.8, respectively. The steps sizes of all the diffusion algorithms are 6.1 and 2.5 for the tonal signal and broadband noise, respectively. The decentralized algorithm uses the same step size as the diffusion algorithm so as to verify the improvement in stability. The VSR parameters $(\kappa_j, \beta_j, \lambda_j(n=0))$ [20] for the tonal signal and broadband noise are (0.01, 0.02, 0) and (0.01, 0.001, 0), respectively. The normalized residual noise is used to evaluate the noise control performance, which is defined as

$$\tau(n) = 10\lg \frac{\sum_{j=0}^{10}|e_j(n)|^2}{\sum_{j=0}^{10}|d_j(n)|^2} \quad (17)$$

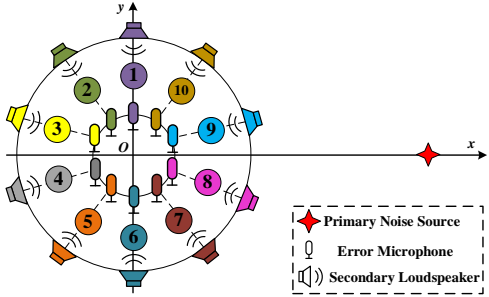

Fig. 3. The simulation setup for 10-channel ANC system.

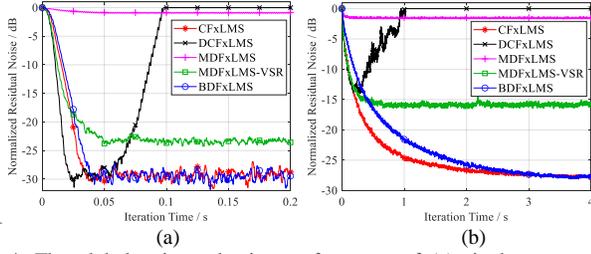

Fig. 4. The global noise reduction performance of (a) single-tone and (b) broadband noise using different algorithms.

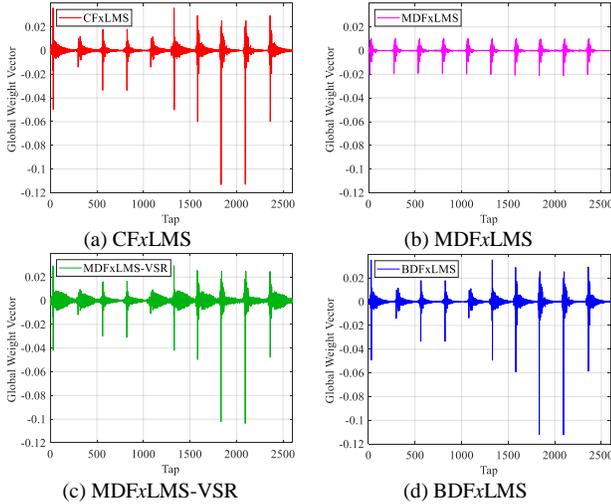

Fig. 5. The global control filter under broadband noise of (a) CF$x$LMS, (b) MDF$x$LMS, (c) MDF$x$LMS based on the VSR and (d) BDF$x$LMS algorithms.

All simulation results for the broadband noise are averaged over 1000 Monte Carlo runs and the results for tonal signals are smoothed with a moving-average window of 10 samples.

The normalized residual noise for the tonal signal and broadband noise are compared in Fig. 4 for different algorithms. It is clear that the CF$x$LMS algorithm achieves the best performance while the DCF$x$LMS algorithm diverges due to the inherent instability, as expected. The MDF$x$LMS algorithm is stable but shows the worst noise reduction performance, which is mainly because all the control filters tend to be consistent in the steady state due to the forced spatial smoothing [20, 21]. This can be verified by comparing the converged solutions of the CF$x$LMS and MDF$x$LMS algorithms for the broadband signal in Figs. 5(a) and (b), respectively. It is clear that the MDF$x$LMS algorithm forces the control filters of different nodes to be the same, which deviates from the converged solution of the centralized algorithm. The VSR strategy reduces the normalized residual noise and the bias compared to the original MDF$x$LMS algorithm, as shown in Figs. 4 and 5(c), but it still cannot converge to the centralized solution.

By contrast, the proposed BDF$x$LMS-BC algorithm completely eliminates the bias introduced by the spatial regularization and converges to the centralized solution. Fig. 4 shows that the BDF$x$LMS-BC algorithm achieves the same noise reduction performance as the CF$x$LMS and outperforms the existing decentralized and distributed algorithms. The normalized residual noise for the tonal signal in Fig. 4(a) is reduced to -30 dB, which is the lower limit because the SNR of the error microphones is set to 30 dB. By comparing Figs. 5(a) and 5(d), it can be seen that the control filters estimated by the BDF$x$LMS-BC algorithm are the same as that by the CF$x$LMS algorithm.

The above results demonstrate that the proposed BDF$x$LMS-BC algorithm based on the neighborhood-wide adaptation and node-specific combination strategy overcomes the instability issues of the DCF$x$LMS algorithm and the bias problems of the MDF$x$LMS algorithms. It should be noted that these improvements come at the expense of higher computational burden, because each node estimates the block vector rather than the node's own control filter. The overall computational complexities of the above algorithms are compared in Table. I for the simulated 10-channel ANC system. It is clear that the proposed BDF$x$LMS-BC achieves better noise reduction performance than the MDF$x$LMS-VSR algorithm with lower computational burden.

TABLE I
THE GLOBAL COMPUTATIONAL COMPLEXITY

| Algorithms | Multiplications | Additions |
|---|---|---|
| CF$x$LMS | 63600 | 60790 |
| DCF$x$LMS | 8450 | 8410 |
| MDF$x$LMS | 17290 | 14650 |
| MDF$x$LMS-VSR | 23628 | 20940 |
| BDF$x$LMS-BC | 22466 | 19812 |

## V. CONCLUSION

In this paper, a BDF$x$LMS-BC algorithm based on the neighborhood-wide adaptation and node-specific combination is proposed. A 10-channel ANC system was simulated to compare the proposed algorithm with the existing decentralized and distributed algorithms. Simulation results show that the proposed BDF$x$LMS-BC obtains consistent noise reduction performance with the centralized algorithm, overcomes the instability risk of the decentralized scheme, and eliminates the estimation bias of the distributed algorithm based on diffusion adaptation. Future work will verify the simulation results through real-time experiments and explore algorithms with further reduced computational complexity.